\documentclass{elsart}

\usepackage{natbib}
\usepackage{graphicx}
\usepackage{amssymb,amsmath} 
\usepackage{graphicx}
\usepackage{ulem}
\usepackage{units} 

\providecommand{\martin}[1]{{#1}}

\begin{document}
\begin{frontmatter}
\title{Three-phase traffic theory and two-phase 
models with a fundamental diagram in the light of empirical stylized facts}

\author[TUD]{Martin Treiber\corauthref{cor1}}, \ead[url]{http://www.mtreiber.de} \ead{treiber@vwi.tu-dresden.de} 
\author[TUD]{Arne Kesting}, 
\author[ETH,HUNG]{and Dirk Helbing} 

\address[TUD]{Technische Universit\"at Dresden, Institute for Transport \& Economics,\\ W\"urzburger Str. 35, D-01187 Dresden, Germany}
\address[ETH]{ETH Z\"urich, UNO D11, Universit\"atsstr. 41, CH-8092 Z\"urich, Switzerland}
\address[HUNG]{Collegium Budapest -- Institute for Advanced Study,\\ Szenth\'aroms\'ag u. 2, H-1014 Budapest, Hungary}
\corauth[cor1]{Corresponding author. Tel.: +49 351 463 36794; fax: +49 351 463 36809}

\begin{abstract}
Despite the availability of large empirical data sets and the
long history of traffic modeling, the theory of traffic congestion on
freeways is still highly controversial. In this contribution, we
compare Kerner's three-phase traffic theory with the phase diagram
approach for traffic models with a fundamental diagram. We discuss the
inconsistent use of the term ``traffic phase'' and show that patterns
demanded by three-phase traffic theory can be reproduced with simple
two-phase models, if the model parameters are suitably specified and
factors characteristic for real traffic flows are considered, such as
effects of noise or heterogeneity or the actual freeway design
(e.g. combinations of off- and on-ramps). Conversely, we demonstrate
that models created to reproduce three-phase traffic theory create
similar spatiotemporal traffic states and associated phase diagrams,
no matter whether the parameters imply a fundamental diagram in
equilibrium or non-unique flow-{\it density} relationships. In conclusion, there are different ways of reproducing the empirical stylized facts of spatiotemporal congestion patterns summarized in this contribution, and it appears possible to overcome the controversy by a more precise definition of the scientific terms and a more careful comparison of models and data, considering effects of the measurement process and the right level of detail in the traffic model used.
\end{abstract}


\end{frontmatter}

\section{Introduction}
The observed complexity of congested traffic flows has puzzled traffic
modelers for a long time (see \cite{Helb-Opus} for an
overview). The most
controversial open problems concern the issue of
faster-than-vehicle characteristic propagation speeds  \citep{Daganzo,Rascle}
and the question whether traffic models with or without a fundamental
diagram (i.e. with or without a unique equilibrium flow-density or
speed-distance relationship) would describe empirical observations
best. While the first issue has been intensively debated recently (see
\cite{HelJoh}, and references therein), this paper addresses the second issue.  
\par
The most prominent approach regarding models {\it without} a
fundamental diagram is the three-phase traffic theory
by~\citep{Kerner-book}. The three phases of this theory are ``free
traffic'', ``wide moving jams'', and ``synchronized flow''. While a
characteristic feature of ``synchronized flow'' is the wide scattering
of flow-density data~\citep{Kerner-Rehb96-2}, many microscopic and
macroscopic traffic models neglect noise effects and the
heterogeneity of driver-vehicle units for the sake of simplicity, and
they possess a unique flow-density or speed-distance relationship
under stationary and spatially homogeneous equilibrium 
conditions. Therefore, Appendix \ref{SCA} discusses some issues concerning 
the wide scattering of  congested traffic flows and how it can be 
treated within the framework of such models.
\par
For models with a fundamental diagram, a phase diagram approach has
been developed~\citep{Phase} to represent the conditions under which
certain traffic states can exist. A favourable property of this
approach is the possibility to semi-quantitatively derive the
conditions for the occurence of the different traffic states from the
instability properties of the model under consideration and the
outflow from congested traffic~\citep{Helb-Phases-EPJB-09}. The phase diagram approach for
models with a fundamental diagram has recently been backed up by
empirical studies~\citep{Schoenhof-TRB09}. Nevertheless, the approach
has been
criticized~\citep{Ker02_PRE,kerner_jpa_2008}, which applies to the
alternative three-phase traffic theory as
well~\citep{Martin-empStates,Schoenhof-TRB09}. While both theories
claim to be able to explain the empirical data, particularly the
different traffic states and the transitions between them, the main
dispute concerns the following points: 
\begin{itemize}
\item Both approaches use an inconsistent terminology regarding the definition of traffic phases and the naming of the traffic states.
\item Both modeling approaches make simplifications, but are confronted with empirical details they were not intended to reproduce (e.g. effects of details of the freeway design, or the heterogeneity of driver-vehicle units).
\item three-phase traffic theory is criticized for being complex,
inaccurate, and inconsistent, and related models are criticized to
contain too many parameters to be 
meaningful~\citep{Helb-crit,Martin-empStates}.
\item It is claimed that the phase diagram of models with a
fundamental diagram would not represent the empirical observed traffic
states and transitions well~\citep{Kerner-book}. In particular, the
``general pattern'' (GP) and the ``widening synchronized pattern''
(WSP) would be missing. Moreover, wide moving jams should always be
part of a ``general pattern'', and homogeneous traffic flows should
not occur for extreme, but rather for small bottleneck strengths.  
\end{itemize}  
In the following chapters, we will try to overcome these problems. In
Sec.~\ref{sec:phenomen} we will 
summarize the stylized empirical facts that are observed on freeways
in many different countries and have to be explained by realistic
traffic models. Afterwards, we will discuss and clarify the concept 
of traffic phases in Sec. \ref{sec:defPhases}. In
Sec.~\ref{sec:phase}, we show that the traffic patterns of three-phase
traffic theory can be simulated by a variety of microscopic and
macroscopic traffic models with a fundamental diagram, if the model
parameters are suitably chosen. For these model parameters, the
resulting traffic patterns look surprisingly similar to simulation
results for models representing three-phase traffic theory, which have
a much higher degree of complexity. Depending on the interest of the
reader, he/she may jump directly to the section of interest. 
Finally, in Sec.~\ref{sec:conclusions}, we will summarize and discuss
the alternative explanation 
mechanisms, pointing out possible ways of resolving the controversy.

\section{\label{sec:phenomen}Overview of 
empirical observations}

In this section, we will pursue a data-oriented approach. Whenever
possible, we describe the observed data without using technical terms used within the
framework of three-phase traffic theory or 
models with a fundamental diagram. In order to show that the following observations are generally valid, we present data from several freeways in
Germany, not only from the German freeway A5, which has been
extensively studied
before~\citep{KernerPinch,KeRe96,Martin-empStates,Schoenhof-TRB09,Bertini-TRB-04,lindgren-A5-trr06}.
Our data from a variety of other countries confirm these observations 
as well~\citep{Zielke-intlComparison}.

\subsection{\label{sec:measurement}Data issues}

In order to eliminate confusion arising from different
interpretations of the data and to facilitate a direct comparison between computer simulations
and observations, one has to simulate the method of data acquisition and the subsequent
processing or interpretation steps as well. We will
restrict ourselves here to the consideration to aggregated stationary detector data
which currently is the main data source of freeway traffic studies. When
comparing empirical and simulation data,  we will focus on the velocity~$V$
(and not the density), since it can be measured directly. In addition
to the aggregation over one-minute time intervals,
we will also aggregate over the freeway lanes. 
This is justified due to the typical synchronization of velocities among 
freeway lanes in all types of congested traffic~\citep{Helb-crit}.

To simulate the measurement and interpretation process, we use
``virtual detectors'' recording the passage time and velocity of each
vehicle. For each aggregation time
interval (typically \unit[60]{s}), we determine 
the traffic flow~$Q$ as the vehicle count divided by the
aggregation time, and the velocity~$V$ as the arithmetic mean value of the
individual vehicles passing in this time period. 
Notice that the arithmetic mean value leads to a systematic
overestimation of velocities in congested situations and that there exist better averaging methods
such as  the harmonic mean~\citep{Opus}.
 Nevertheless, we will use the above procedure because this is
the way in which empirical data are typically evaluated by detectors. 

Since freeway detectors are positioned only at a number of discrete
locations, interpolation techniques have to be applied to
reconstruct the observed spatiotemporal dynamics at any point in a given spatiotemporal
region. If the detector locations are not further apart than about \unit[1]{km}, it
is sufficient to apply a linear smoothing/interpolating filter, or
even to plot the time series of the single detectors in a suitable
way (see, e.g. Fig.~1 in~\cite{Martin-empStates}). This condition,
however, severely restricts the selection of suitable freeway
sections, which is one of the reasons why empirical traffic studies in Germany
have been concentrated on a \unit[30]{km} long section of the
Autobahn A5 near Frankfurt.  For most other freeway sections showing
recurrent congestion patterns, two neighboring detectors are
\unit[1-3]{km} apart, which is of the same order of magnitude as typical wavelengths
of non-homogeneous congestion patterns and therefore leads to
ambiguities as demonstrated by~\cite{Treiber-smooth}. Furthermore, the
heterogeneity of traffic flows and measurement noise lead to
fluctuations obscuring the underlying patterns. 

Both problems can be overcome by post-processing the aggregated detector 
data~\citep{CasWin95,coifmanSpatiotemp,Bertini-TRB-04,CTDMunozDaganz-x,Schreck-ASM,Treiber-smooth}.
Furthermore, \cite{KerASDA-tec} have proposed a method called ``ASDA/FOTO'' for
short-term traffic prediction. Most of these methods, however, cannot be applied for the present
investigation since they do not provide continuous velocity
estimates for all points $(x,t)$ of a certain spatiotemporal region,
or because they are explicitly
based on models. (The method ASDA/FOTO, for example, is based on 
three-phase traffic theory.) We will therefore use the adaptive
smoothing method~\citep{Treiber-smooth}, which has recently been validated
with empirical data of very high spatial resolution \citep{ASM-CACAIE}. 
In order to be consistent, we will apply this method 
to both, the real data and the virtual detector data of our computer simulations.

\subsection{\label{sec:3ddata}Spatiotemporal data}

In this section, we will summarize the \textit{stylized
facts} of the spatiotemporal evolution of congested
traffic patterns, i.e., typical empirical findings that are 
persistently observed on various freeways all over the world.  
In order to provide  a comprehensive list 
as a \textit{testbed} for traffic models and theories, we will summarize below all relevant
findings, including already published ones:  

\begin{figure}
\centering
\includegraphics[width=1.0\textwidth]{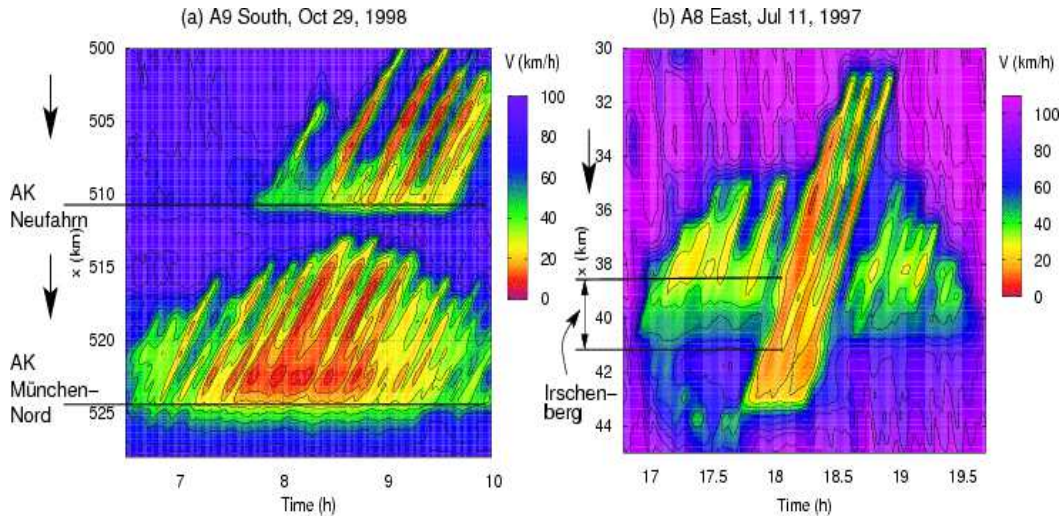}

 \caption{\label{fig:empdata1}Spatiotemporal dynamics of the average
 velocity on two different freeways. (a) German freeway A9 in direction South, located in the area North of
 Munich. Horizontal lines indicate two intersections (labelled ``AK''), which cause bottlenecks, since they consume some of the freeway capacity. The traffic direction is shown by arrows. (b) German freeway A8 in direction East, located about~\unit[40]{km} East of Munich. Here, the bottlenecks are caused by uphill and downhill gradients around ``Irschenberg'' and by an accident at
 $x=\unit[43.5]{km}$ in the time period between \unit[17:40]{h} and
 \unit[18:20]{h}.}
\end{figure}
\begin{figure}
\centering
\includegraphics[width=1.0\textwidth]{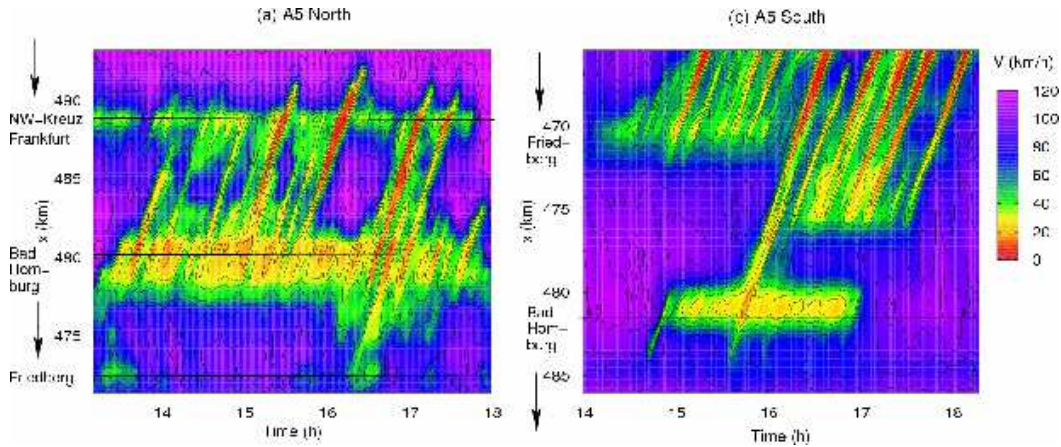}

 \caption{\label{fig:empdata2}Spatiotemporal velocity profiles for the
 German freeway A5 North near Frankfurt/Main (both directions). Arrows indicate the direction of travel.}

\end{figure}

\begin{enumerate}
\item \textit{Congestion patterns on real (non-circular) freeways 
are typically caused by bottlenecks in combination with a perturbation in the
traffic flow.} An extensive study of the breakdown
phenomena on the German freeways A5-North 
and A5-South by~\cite{Martin-empStates}, 
analyzing about 400 congestion patterns, did not find examples where 
there was an apparent lack of a bottleneck.  This is in agreement with
former investigations of the Dutch freeway~A9, the
German freeway A8-East and West, and the German freeway A9-South~\citep{Opus}.  
Nevertheless, it may {\it appear} to drivers entering a traffic jams on a 
homogeneous freeway section that they are experiencing a "phantom traffic jam", 
i.e. a traffic jam without any apparent reason. In these cases, however, the triggering 
bottleneck, which is actually the reason for the traffic jam, is
located downstream,  
potentially in a large distance from the driver location (see Fig.~13  of~\cite{Martin-empStates} 
or in Fig.~1(a) of~\cite{Helb-Phases-EPJB-09}).

\item \textit{The bottleneck may be caused by various reasons} such as 
isolated on-ramps or off-ramps, combinations thereof such as junctions or
intersections (Fig.~\ref{fig:empdata1}(a) and~\ref{fig:empdata2}),
local narrowings or reductions of the number of lanes, accidents,
or gradients. As an example, Fig.~\ref{fig:empdata1}(b) shows a composite
congestion pattern on the German freeway A8-East caused by uphill and downhill
gradients (``Irschenberg'') in the region $\unit[38]{km} \le x \le
\unit[41]{km}$, and an additional obstruction by an accident at
$x=\unit[43.5]{km}$ in the time period between \unit[17:40]{h} and \unit[18:15]{h}.

\item \textit{The congestion pattern is either localized with
a constant width of the order of \unit[1]{km}, or it is spatially
extended with a time-dependent  
extension.}  
Localized congestion patterns either remain stationary
at the bottleneck, or they move upstream at a characteristic speed
$c_{\rm cong}$. Typical values of $c_{\rm cong}$ are between $\unit[-20]{km/h}$ and
$\unit[-15]{km/h}$, depending on the country and
traffic composition \citep{Zielke-intlComparison}, but not on the  type of
congestion.  About 200 out of 400 breakdowns observed by~\cite{Martin-empStates} correspond to extended patterns.

\item \textit{The downstream front of congested traffic is either fixed at the
bottleneck, or it moves upstream with the characteristic speed
$c_{\rm cong}$}~\citep{Helb-Phases-EPJB-09}. Both, fixed and moving downstream fronts can 
occur within one and the same congestion pattern. This can be seen in 
Fig.~\ref{fig:empdata1}(a), where the stationary downstream congestion front at $x=\unit[476]{km}$ (the location of the temporary bottleneck caused by an incident) starts moving upstream at \unit[17:30]{h}. Such a ``detachment'' of the downstream congestion front occurs, for example, when an accident site has been cleared, and it is one of two ways in which the dissolution of traffic congestion starts (see next item for the second one).

\item \textit{The upstream front of spatially extended congestion patterns 
has no characteristic speed.} Depending on the traffic demand and
the bottleneck capacity, it can propagate upstream (if the demand exceeds the capacity)
or downstream (if the demand is below
capacity)~\citep{Helb-Phases-EPJB-09}. 
This can be seen in all extended congestion patterns of Fig.~\ref{fig:empdata1}
(see also~\cite{Schoenhof-TRB09,Kerner-book}).
The downstream movement of the congestion front towards the bottleneck is the second and most frequent way in which congestion patterns may dissolve.

\item \textit{Most extended traffic
 patterns show some  ``internal structure'' propagating upstream approximately at 
the same characteristic speed $c_{\rm cong}$.}
Consequently, all spatiotemporal structures in
Figs.~\ref{fig:empdata1} and~\ref{fig:empdata2} (sometimes termed
``oscillations'', ``stop-and-go traffic'', or ``small 
jams''), move in parallel~\citep{DaganzoLong1999,Mauch-Cassidy,Zielke-intlComparison}.

\item \textit{The periods and wavelengths of internal structures in congested traffic states tend to decrease as the severity of congestion increases.} This applies in particular to measurements of the average velocity. (See, for example, Fig.~\ref{fig:empdata1}(a), where the greater of two bottlenecks, located
at the Intersection M\"unchen-Nord, produces oscillations of a higher
frequency. Typical periods  of the internal
quasi-periodic oscillations vary between about \unit[4]{min} and
\unit[60]{min}, corresponding to wavelengths between \unit[1]{km}
and \unit[15]{km}~\citep{Helb-crit}.

\item \textit{For bottlenecks of moderate strength, 
the amplitude of the internal structures tends to increase while
propagating upstream}. This can be seen in \textit{all} empirical traffic states shown in this
contribution, and also in~\cite{Schoenhof-TRB09,Helb-Phases-EPJB-09}. It can also
be seen in the corresponding velocity time series, such as the ones in
Fig.~12 of~\cite{Opus}, in~\cite{Zielke-intlComparison},
or in \textit{all} relevant time series shown
in Chapters~9-13 of~\cite{Kerner-book}.
The oscillations may already be visible at the downstream boundary
(Fig.~\ref{fig:empdata1}(b)), or emerge further upstream
(Figs.~\ref{fig:empdata1}(a), \ref{fig:empdata2}(a)). During their
growth, neighboring perturbations may merge (Fig.~1
in~\cite{Schoenhof-TRB09}), or propagate unaffected (Fig. 1). At the upstream
end of the congested area, the oscillations may eventually become 
isolated ``wide jams'' (Fig.~\ref{fig:empdata2}) or remain part of a
compact congestion pattern (Fig.~\ref{fig:empdata1}).

\item \textit{Light or very strong bottlenecks may cause extended
traffic patterns, which appear homogeneous (uniform in space),} see,
for example, Figs. 1(d) and 1(f) of ~\cite{Helb-Phases-EPJB-09}. 
Note however that, for strong bottlenecks (typically caused by accidents), 
the empirical evidence has been controversially debated, in particular as 
the oscillation periods at high densities reach the same order of magnitude as the smoothing time
window that has typically been used in previous studies (cf. point 7 above). This makes 
oscillations hardly distinguishable from noise.\footnote{Moreover, speed variations between 'stop and
slow' may result from problems in maintaining low speeds (the gas
and brake pedals are difficult to control in this regime), and thus
are different from the collective dynamics at higher speeds. In any
case, this is not a crucial point since there are models that 
can be calibrated to generate homogeneous patterns for high bottleneck
strengths (restabilization), or not, see Eq.~(\ref{IDM-HCT}) in
Sec.~\ref{sec:simIDM} below.} See Appendix \ref{APP} for a further discussion of this 
issue.

\end{enumerate}

Note that the above stylized facts have not only be observed in Germany, but also 
in other countries, e.g. the USA, Great Britain, and the Netherlands
\citep{Zielke-intlComparison,Helb-Phases-EPJB-09,Wilson-Pattern2008,ASM-CACAIE}.
Furthermore, we find that many congestion patterns are composed of several of the 
elementary patterns listed above \citep{Martin-empStates}. For example, the congestion pattern observed in
 Fig. \ref{fig:empdata2}(b) can be decomposed into moving
and stationary localized patterns as well as extended
patterns.

The source of probably most controversies in traffic theory is an
observed spatiotemporal structure called the \textit{``pinch effect''} or
\textit{``general pattern''}~\citep{Kerner-Rehb96-2}, see \cite{Kerner-book} for details and Fig.~1
of~\cite{Schoenhof-TRB09} for a typical example of the spatiotemporal evolution. 
From the perspective of the above list, 
this pattern relates to {\it stylized facts 6 and 8}, i.e., it  has the following
features: (i) relatively stationary congested traffic (\textit{pinch
region}) near the downstream front, (ii) small perturbations that grow
to oscillatory structures as they travel further upstream, (iii) some
of these structures grow to form ``wide jams'', thereby suppressing
other small jams, which either merge or dissolve. The question is whether this congestion
pattern is composed of several elementary congestion patterns or a separate, elementary
pattern, which is sometimes called ``general
pattern''~\citep{Kerner-book}. This will be addressed in
Sec.~\ref{sec:pinch}.

\section{\label{sec:defPhases}The meaning of traffic phases}

The concept of ``phases'' has originally been used in areas such as thermodynamics,
physics, and chemistry. In these systems, ``phases'' mean different aggregate states
(such as solid, fluid, or gaseous; or different material compositions in metallurgy; or
different collective states in solid state physics).  When certain 
``control parameters'' such as the pressure or temperature in the system are changed,
the aggregate state may change as well, i.e. a qualitatively different macroscopic 
organization of the system may result. If the transition is abrupt, one speaks of first-order
(or ``hysteretic'', history-dependent) phase transitions. Otherwise, if the transition is continuous,
one speaks of second-order phase transitions.\footnote{In order to measure whether a phase
transition is continous or not, a suitable ``order parameter'' needs to be defined and measured.}
\par
In an abstract space, whose axes are defined by the control parameters, it is useful
to mark parameter combinations, for which a phase transition occurs, by 
lines or ``critical points''. Such illustrations
are called phase diagrams, as they specify the conditions, 
under which certain phases occur. 
\par
Most of the time, the terms ``phase'' and ``phase diagram'' are applied to large (quasi-infinite), spatially closed, and homogeneous systems in thermodynamic equilibrium, where the phase can be determined in any point of the system. When transferring these concepts to traffic flows,
researchers have distinguished between one-phase, two-phase, and
three-phase models. The number of phases is basically related to the
(in) stability properties of the traffic flows (i.e. the number of
states that the instability diagram distinguishes). The equilibrium
state of one-phase models is a spatially homogeneous traffic state
(assuming a long circular road without any bottleneck). An example
would be the  Burgers equation~\citep{Whitham-waves}, i.e. a
Lighthill--Whitham--Richard model~\citep{Lighthill-W,Richards} with
diffusion term. Two-phase models would additionally produce
oscillatory traffic states such as wide moving jams or stop-and-go
waves, i.e. they require some instability mechanism~\citep{wagner_Phases2008}.Three-phase models introduce another traffic state,
so-called ``synchronized flow'', which is characterized by a
self-generated scattering of the traffic variables. It is not clear,
however, whether this state exists in reality in the absence of
spatial inhomogeneities (freeway bottlenecks).\footnote{In fact, it
even remains to be shown whether Kerner's ``three-phase''
car-following models~\citep{Kerner-Mic,kerner_jpa_2006} or other 
three-phase models really have three phases in the thermodynamic sense pursued by
\cite{wagner_Phases2008}.}
\par
Note, however, that the concept of phase transitions has also been
transferred to non-equilibrium systems, i.e. driven, open systems with
a permanent inflow or outflow of energy, inhomogeneities, etc. 
This use is common in systems theory. For
example, one has introduced the concept of boundary-induced phase
transitions~\citep{KrugLetter,Popkov-boundaryPhase,Appert-boundaryPhase}.
From this perspective, the Burgers
equation can show a boundary-induced phase transition from a free-flow
state with {\it forwardly} propagating congestion fronts to a
congested state with {\it upstream} moving perturbations of the
traffic flow. This implies that the Burgers equation (with one
equilibrium phase) has {\it two non-}equilibrium phases. Analogously,
two-phase models (in the previously discussed, thermodynamic sense)
can have more than two {\it non-}equilibrium phases. However, to avoid
confusion, one often uses the terms ``(spatiotemporal) traffic
patterns'' or ``(elementary) traffic states'' rather than
``non-equilibrium phases''. For example, the gas-kinetic-based traffic
model (GKT model) or the intelligent driver model (IDM), which are
two-phase models according to the above classification, may display
{\it several} congested traffic states besides free traffic flow~\citep{Opus}.
The phase diagram approach to traffic modeling 
proposed by~\cite{Phase} was originally presented for an open traffic system
with an on-ramp. It shows the qualitatively different, spatiotemporal
traffic patterns as a function of the freeway flow and the bottleneck
strength. 

Note, however, that the resulting traffic state may depend on the history (e.g. the size of perturbations in the traffic flow), if traffic flows have the property of metastability. 
\par
The concept of the phase diagram has been taken up by many authors and
applied to the spatiotemporal traffic patterns (non-equilibrium
phases) produced in many models 
\citep{Lee,Lee99,Kerner-book,siebel2006}. Besides on-ramp scenarios,
one may study scenarios with flow-conserving bottlenecks (such as lane
closures or gradients) or with combinations of several bottlenecks. It
appears, however, that the traffic patterns for freeway designs with
several bottlenecks can be understood, based on the {\it combination}
of elementary traffic patterns appearing in a system with a {\it
single} bottleneck and interaction effects between these patterns
\citep{Martin-empStates,Helb-Phases-EPJB-09}
\par
The resulting traffic patterns as a function of the flow conditions
and bottleneck strengths (freeway design), and therefore the
appearance of the phase diagram, depend on the traffic model and the
parameters chosen. Therefore, the phase diagram approach can be used
to classify the large number of traffic models into a few
classes. Models with qualitatively similar phase diagrams would be
considered equivalent, while models producing different kinds of
traffic states would belong to different classes. The grand challenge
of traffic theory is therefore to find a model and/or model
parameters, for which the congestion patterns match the stylized facts
(see Sec. \ref{sec:3ddata}) and for which the phase diagram agrees with the
empirical one \citep{Martin-empStates,Helb-Phases-EPJB-09}. This issue will
be addressed in Sec.~\ref{sec:phase}
\par
For the understanding of traffic dynamics one may ask which of the
two competing phase definitions (the thermodynamic or the non-equilibrium one) 
would be more relevant for observable phenomena. Considering the
stylized facts (see Sec.~\ref{sec:phenomen}),
it is obvious that boundary conditions and inhomogeneities play an important role for the resulting traffic patterns. This clearly favours the dynamic-phase concept over the definition
of thermodynamic equilibrium phases: Traffic patterns are easily observable and also
relevant for applications. (For calculating traveling times, one needs
the spatiotemporal dynamics of the traffic pattern, and not the
thermodynamic traffic phase.) Moreover, thermodynamic phases are not
\textit{observable} in the strict sense, because real traffic systems are not quasi-infinite, homogeneous, closed systems. Consequently, when assessing the quality of a given model,
it is of little relevance whether it has two or three physical phases, as long as it
correctly predicts the observed spatiotemporal patterns,
including the correct conditions for their
occurrence. Nevertheless, the thermodynamic phase concept (the instability diagram)
is relevant for \textit{explaining} the mechanisms leading to the different patterns.
In fact, for models with a fundamental diagram, it is possible to
derive the phase diagram of traffic states from the instability
diagram, if bottleneck effects and the outflow from congested traffic
are additionally considered~\citep{Phase}.

\section{\label{sec:phase}Simulating the spatiotemporal traffic dynamics}

In the following, we will show for specific traffic models that not only three-phase traffic theory,
but also the conceptionally simpler two-phase models (as introduced in
Sec.~\ref{sec:defPhases}) can display all stylized facts mentioned in Sec.
\ref{sec:phenomen}, if the model parameters are suitably chosen. 
This is also true for patterns that were attributed
exclusively to three-phase traffic theory such as the pinch effect or the widening 
synchronized pattern (WSP). 

Considering the dynamic-phase definition of Sec.~\ref{sec:defPhases},
the simplest system that allows to reproduce realistic congestion
patterns is an open system with a bottleneck. When simulating an
on-ramp bottleneck, the possible flow conditions can be characterized by the
{\it upstream} freeway flow (``main inflow'') and the ramp flow, considering the number 
of lanes~\citep{Phase}. The {\it downstream} traffic flow under free and congested 
conditions can be determined from these quantities. 
When simulating a flow-conserving (ramp-less)
bottleneck, the ramp flow is replaced by the \textit{bottleneck
strength} quantifying the degree of local capacity
reduction~\citep{Treiber-TGF99}. 

Since many models show hysteresis effects, i.e. discontinuous, history-dependent 
transitions, the time-dependent traffic conditions before the onset of congestion are 
relevant as well. In the simplest case, the response of the system is tested (i) for minimum
perturbations, e.g. slowly increasing inflows and ramp flows, and
(ii) for a large perturbation. The second case is usually studied by
generating a wide moving jam, which can be done by temporarily blocking the outflow. 
Additionally, the model parameters characterizing the bottleneck situation
have to be systematically varied and scanned through. This is, of course, a
time-consuming task since producing a single point in this multi-dimensional 
space requires a complete simulation run (or even to average over several
simulation runs).

\subsection{Two-phase models}
%
\cite{wagner_Phases2008}
classify models with a fundamental diagram that show
dynamic traffic instabilities in a certain density range, as two-phase models. 
Alternatively, these models are referred to as ``models
within the fundamental diagram approach''. Note, however, that
certain models with a unique fundamental diagram are {\it one}-phase models
(such as the Burgers equation).
Moreover, some models such as the KK model
can show one-phase, two-phase or three-phase behavior,
depending on the choice of model parameters (see Sec.~\ref{sec:three}).

A microscopic two-phase model necessarily has a dynamic
acceleration equation or  contains time delays such as a reaction
time. For macroscopic models, a necessary (but not sufficient)
condition for two phases is that the model contains a dynamical equation for the
macroscopic velocity.

\subsubsection{Traffic patterns for a macroscopic traffic model}
%
We start with results for the gas-kinetic-based traffic model \citep{helbing1996gas,GKT}.  
Like other macroscopic traffic models, the GKT
model describes the dynamics of aggregate quantities, but besides the 
vehicle density $\rho(x,t)$ and average velocity $V(x,t)$, it also considers the velocity
variance $\theta(x,t)=A(\rho(x,t))V(x,t)^2$ as a function of
velocity and density.  

The GKT model has five parameters $v_0$, $T$, $\tau$, $\gamma$, and
$\rho_{\rm max}$ characterizing the driver-vehicle units, see
Table~\ref{tab:GKT}. In contrast to other popular second-order
models~\citep{Payne-book,KK-93,Lee99,hoogendoorn2000-multiclass}, the
GKT model distinguishes between the desired time gap $T$ when
following other vehicles, and the much larger acceleration time $\tau$
to reach a certain desired velocity. Furthermore, the drivers of the
GKT model ``look ahead'' by a certain multiple $\gamma$ of the
distance to the next vehicle.  The GKT model also contains a
variance function $A(\rho)$ reflecting statistical properties of the traffic data.
Its form can be empirically determined (see Table~\ref{tab:GKT}). For the 
GKT model equations, we refer to~\cite{GKT}.

\begin{table}

\caption{\label{tab:GKT}The two parameter sets 
for the GKT model~\protect\citep{GKT} used in this paper. The four
last parameters specify the velocity
variance prefactor $A(\rho)= A_0 + (A_{\rm max}-A_0)/2 \{
       \tanh [  (\rho-\rho_{\rm c})/\Delta \rho ]  + 1\}$
}

\begin{center}
\begin{tabular}{lll}
Model parameter & Value set~1 & Value set~2 \\ \hline
Desired velocity $v_0$ & \unit[120]{km/h} & \unit[120]{km/h} \\
Desired time gap $T$ & \unit[1.35]{s} & \unit[1.8]{s} \\
Acceleration time $\tau$ & \unit[20]{s} & \unit[35]{s} \\
Anticipation factor $\gamma$ & 1.1 & 1.0\\
Maximum density $\rho_{\rm max}$ & 140/km & 140/km \\ \hline
Variance prefactor $A_0$ for free traffic & 0.008 & 0.01\\
Variance prefactor $A_{\rm max}$ for congested traffic &
  0.038 & 0.03 \\
Transition density free-congested $\rho_c$ 
  & $0.27 ~ \rho_{\rm max}$ &  $0.4 ~ \rho_{\rm max}$ \\
Transition width $\Delta \rho$ 
  & $0.07 ~ \rho_{\rm max}$ & $0.05 ~ \rho_{\rm max}$
\end{tabular}
\end{center}
\end{table}

\begin{figure}
\centering
\includegraphics[width=1.0\textwidth]{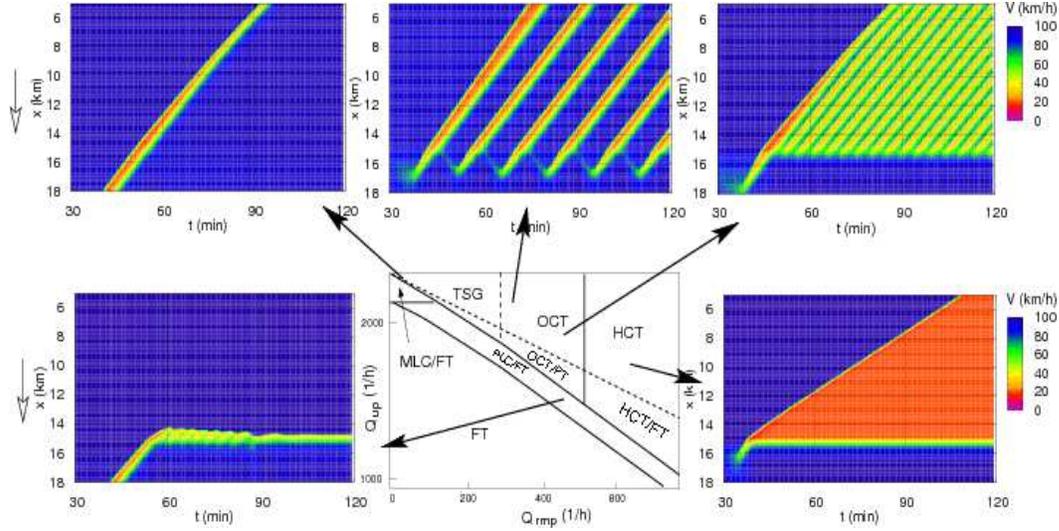}

 \caption{\label{fig:phasediagGKT1b}Congested traffic patterns as a
 function of the dynamic phase space spanned by the main inflow
 and the ramp inflow for the GKT model with parameter set 1 in
 Table~\protect\ref{tab:GKT}. The dotted line indicates the
 maximum traffic demand for which free flow can be sustained.  Below
 this line, congestion patterns can only be triggered by
 perturbations. For this purpose, a moving jam has been generated at
 the downstream boundary in the two plots on the left-hand side. The
 abbreviations denote free traffic (FT), pinned localized cluster
 (PLC), moving localized cluster (MLC), homogeneous congested traffic
 (HCT), oscillatory congested traffic (OCT), and triggered stop-and-go
 (TSG) pattern.}

\end{figure}

We have simulated an open system with an on-ramp as a function of the main
flow and the ramp flow, using the two parameter sets listed in
Table~\ref{tab:GKT}. In contrast to the simulations
in~\cite{Phase}, we added variations of the on-ramp flow
with an amplitude of \unit[20]{vehicles/h} and a mean
value of zero to compensate for the overly
smooth merging dynamics in macroscopic 
models, when mergings are just modeled by constant (or slowly varying)
source terms in the continuity equation.
For parameter set~1, we obtain the results of Fig.~\ref{fig:phasediagGKT1b}, i.e.,
the phase diagram found by~\cite{Phase} and by~\cite{Lee}. It contains five
congested traffic patterns, namely {\it pinned localized clusters (PLC), moving localized clusters (MLC), 
triggered stop-and-go waves (TSG), oscillating congested traffic (OCT), and homogeneous
congested traffic (HCT)}. The
OCT and TSG patterns look somewhat similar, and there is no discontinuous transition
between these patterns. This has been indicated by a dashed instead of
a solid line in the  phase diagram. Furthermore, notice that the two
localized patterns MLC and PLC are only obtained, when sufficiently
strong temporary perturbations occur in addition to the stationary
on-ramp bottleneck. Such perturbations may, for example, result from a temporary
peak in the ramp flow or in the main inflow (which may be caused by forming vehicle platoons, when
slower trucks overtake each other, see \cite{Martin-empStates}). Furthermore, the perturbation may be an upstream
moving traffic jam entering a bottleneck area 
(see Fig.~\ref{fig:empdata2}(b) and \cite{Phase}). This case has been assumed here. 

\begin{figure}
\centering
\includegraphics[width=1.0\textwidth]{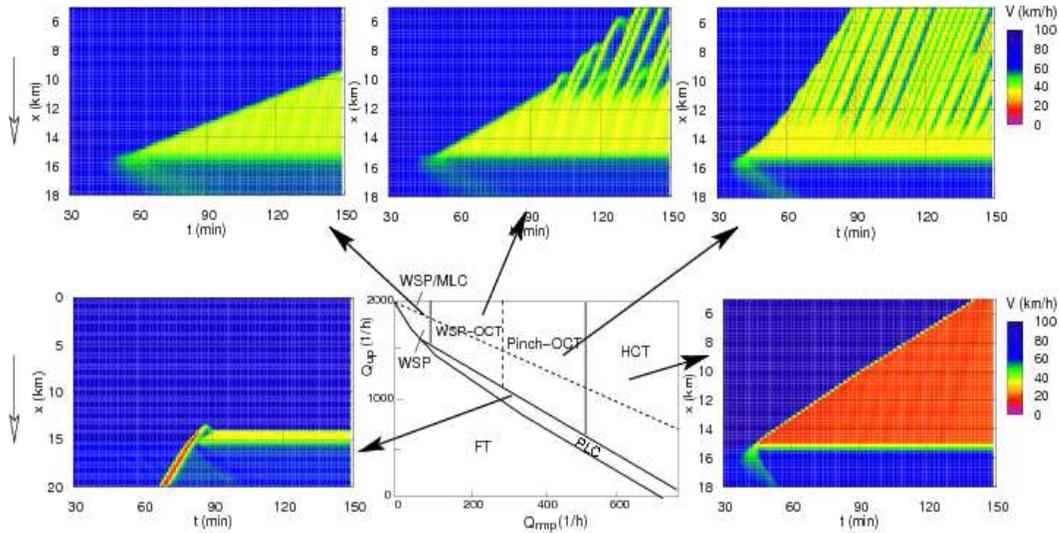}

 \caption{\label{fig:phasediagGKT2b}Dynamic phase diagram of congested
 traffic patterns for the GKT model as in
 Fig.~\protect\ref{fig:phasediagGKT1b}, but this time using parameter
 set~2 from Table~\protect\ref{tab:GKT}. While the TSG pattern is
 missing, the two additional patterns ``pinch region'' and WSP
 (widening synchronized pattern) are produced (see the main text for
 details).}

\end{figure}

When simulating the same system, but this time using parameter set~2
of Table~\ref{tab:GKT}, we obtain the PLC, MLC, OCT and HCT states
as in the first simulation, see Fig.~\ref{fig:phasediagGKT2b} (the MLC
pattern is not shown). However, instead of the TSG state, we find
two new patterns. For very light bottlenecks (small ramp
flows), we observe a light form of homogeneous congested traffic that
has the properties of the \textit{widening synchronized pattern} (WSP)
proposed by~\cite{Kerner-book}. 
Remarkably, this state is stable or metastable, otherwise moving jams should 
emerge from it in the presence of small-amplitude variations of the ramp
flow. Although the WSP-properties
of being extended and homogeneous in space are the same as for the HCT
state, WSP occurs for light bottlenecks,
while HCT requires strong bottlenecks.  Moreover, the two patterns are
separated in the phase diagram by oscillatory states that occur for moderate bottleneck strengths.

The second new traffic pattern is a congested state which consists of a
stationary downstream front at the on-ramp bottleneck, homogeneous, 
light congested traffic near the ramp, and velocity oscillations
(``small jams'' or OCT) further upstream. These are the signatures of
the pinch effect. Similarly to the transition TSG$\leftrightarrow$OCT in the
dynamic phase diagram of Fig.~\ref{fig:phasediagGKT1b}, there is no
sharp transition between light congested traffic  and OCT.

\begin{figure}
\centering
\includegraphics[width=0.9\textwidth]{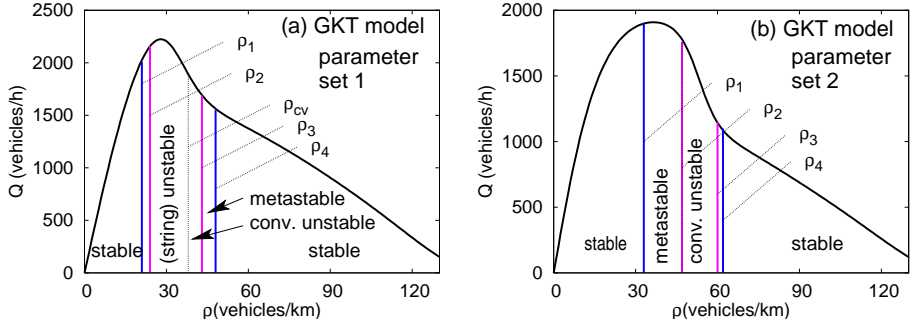}

 \caption{\label{fig:stabdiagGKT}Fundamental diagrams and stability
 regimes of the GKT model for the two parameter sets in
 Table~\protect\ref{tab:GKT}. The so-called {\it critical densities} $\rho_1$,
 $\rho_2$, $\rho_3$, and $\rho_4$ correspond to the densities at which
 the transitions stable$\leftrightarrow$metastable$\leftrightarrow$unstable$\leftrightarrow$metastable$\leftrightarrow$stable occur. For
 $\rho>\rho_{\rm cv}$, the instabilities are of a convective
 type. For the existence of a widening synchronized pattern (WSP), the critical density $\rho_2$
 must be on the ``congested'' side of the fundamental diagram.}

\end{figure}

The corresponding stability diagrams shown in
Fig.~\ref{fig:stabdiagGKT} for the two parameter sets are consistent
with these findings: In contrast to parameter set~1, parameter set~2 leads to a
small density range of metastable (rather than unstable) congested
traffic near the maximum flow, which is necessary for the occurence of the
WSP. Furthermore, parameter set 2 leads to a wide density range of
convectively unstable traffic, which favours the pinch effect as will be
discussed in Sec.~\ref{sec:GP2}.

Finally, we note that the transition from free traffic to extended
congested traffic is of first order. The associated  hysteresis
(capacity drop) is reflected in the phase diagram of Fig. 4 by the
vertical distance between the dotted line and the line separating the
PLC pattern from spatially extended traffic patterns, and also by the large
metastable density regime in the stability diagram (see Fig. 5). 
Note, that the optimal velocity model, in contrast,
behaves nonhysteretic (Kerner and Klenov, 2006, Sec. 6.2), 
which is not true for the microscopic models discussed in the next subsection.

\subsubsection{\label{sec:simIDM}Traffic patterns in microscopic traffic models}
%
In order to investigate the generality of the above results, we have
simulated the same traffic system also with the {\it intelligent driver model (IDM)}
as one representative of two-phase microscopic traffic models with 
continuous dynamics~\citep{Opus}.

The IDM specifies the acceleration $dv_{\alpha}/dt$ of vehicle
$\alpha$ following a leader $\alpha-1$ (with the bumper-to-bumper distance
$s_{\alpha}$ and the relative velocity $\Delta
v_{\alpha}=v_{\alpha}-v_{\alpha-1}$) as a continuous deterministic
function with five model parameters.  The desired velocity $v_0$ and
the time gap $T$ in equilibrium have the same meaning as in the GKT
model. The actual acceleration is limited by the maximum acceleration
$a$. The ``intelligent'' braking strategy generally limits the
decelerations, to the comfortable value $b$, but it allows for higher
decelerations if this is necessary to prevent critical situations or
accidents.  Finally, the gap to the leading vehicle in standing traffic
is represented by $s_0$. Notice that the sum of $s_0$ and the (dynamically
irrelevant) vehicle length  $l_0$ is equivalent to the inverse of the GKT
parameter $\rho_{\rm max}$.

It has been shown that the IDM is able to produce the
five traffic patterns PLC, MLC, TSG, OCT, and HCT found in the GKT model 
with parameter set 1 \citep{Opus}. Here, we want to investigate whether
the IDM can also reproduce the ``new'' patterns shown in 
Fig.~\ref{fig:phasediagGKT2b}, i.e., the WSP and the pinch effect. For this purpose, 
we slightly modify the simulation model as compared to the assumptions made in 
previous publications:

\begin{figure}
\centering
\includegraphics[width=1.0\textwidth]{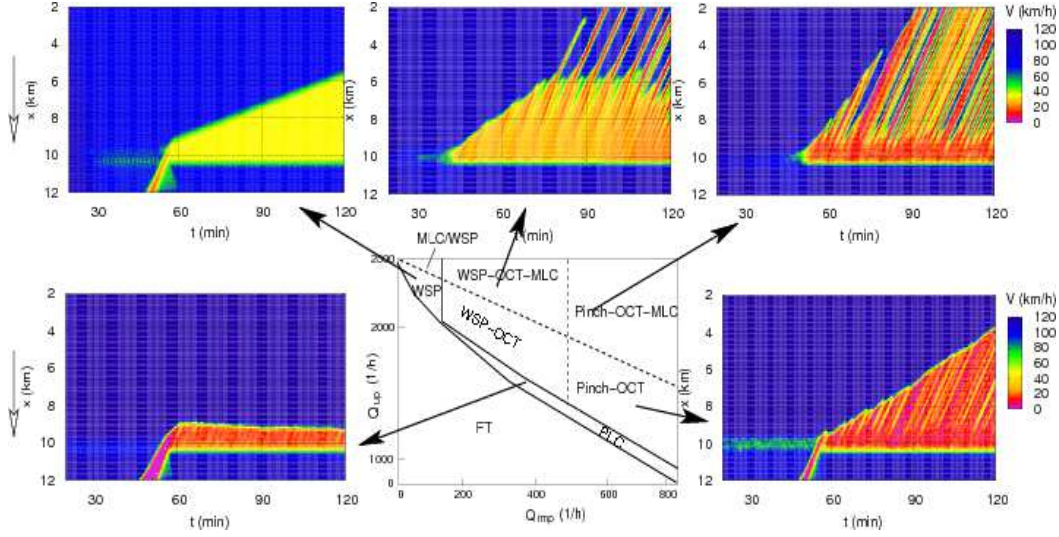}

  \caption{\label{fig:phasediagIDM2a}Dynamic phase diagram of
  on-ramp-induced congested traffic patterns for the Intelligent
  Driver Model with the parameters given in
  Sec.~\protect\ref{sec:simIDM}. As in the previous diagrams, the dashed
  line indicates the maximum traffic volume for which free
  flow can be sustained. In order 
   to trigger the congestion patterns at the bottleneck, \martin{a moving jam
is introduced
 at the downstream boundary for the three simulations corresponding to
points below the dashed line  (metastable
regimes).}
}
\end{figure}

\begin{figure}
\centering
\includegraphics[width=0.45\textwidth]{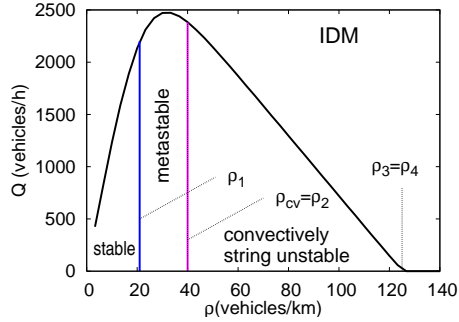}

 \caption{\label{fig:stabdiagIDM}Fundamental diagram and stability
 regions of the IDM for the parameters 
$v_0=\unit[120]{km/h}$,
$T=\unit[1]{s}$, $s_0=\unit[2]{m}$, $a=\unit[1.2]{m/s^2}$, and
$b=\unit[1.3]{m/s^2}$ used in this paper. In
contrast to  the original specifications by~\cite{Opus}, traffic flow
at capacity is metastable rather than linearly unstable here, and the
linear string instability for higher densities is always of the convective type.
See Fig.~\protect\ref{fig:stabdiagGKT} for the definition of the
 critical densities $\rho_i$ and $\rho_{\rm cv}$.}

\end{figure}

\begin{itemize}
\item Instead of a ``flow-conserving bottleneck'' 
we simulate an on-ramp. Since the focus is
not on realistic lane-changing and merging models
we simulate here a main road 
consisting only of one lane and keep the merging rule simple: As soon as an
on-ramp vehicle reaches the merging zone of \unit[600]{m} length, it is
centrally inserted into the largest gap within the on-ramp zone with a
velocity of 60\% of the actual velocity of the leading vehicle on the destination lane.
\item The IDM parameters 
have been changed such that traffic flow at maximum capacity is
{\it meta}stable rather than linearly {\it un}stable. 
This can be reached by increasing the maximum acceleration
$a$. Specifically, we assume $v_0=\unit[120]{km/h}$,
$T=\unit[1]{s}$, $s_0=\unit[2]{m}$, $a=\unit[1.2]{m/s^2}$, and
$b=\unit[1.3]{m/s^2}$. Furthermore, the vehicle length $l_0$ is set
to~\unit[6]{m}.\footnote{This value of $l_0$ is reasonable
for mixed traffic containing a considerable truck fraction.} 
Note that there is actually empirical evidence that
flows are metastable at densities corresponding to capacity
~\citep{helbing2009power,Helb-Phases-EPJB-09}: A growing
vehicle platoon behind overtaking trucks is stable, as long as there
are no significant perturbations in the traffic flow. However, weaving
flows close to ramps can produce perturbations that are large enough
to cause a traffic breakdown, when the platoon reaches the
neighborhood of the ramp. 
\end{itemize}
Figure~\ref{fig:phasediagIDM2a} shows that, with one exception, the
congestion patterns obtained for the IDM model with (meta-)stable maximum flow
are qualitatively the same as for the GKT model with parameter set~2
(see Fig.~\ref{fig:phasediagGKT2b}). 
\martin{As for the GKT model, 
all transitions from free  to congested states are
hysteretic, i.e., the corresponding regions in the phase diagram extend
below the dashed line, where free traffic can be sustained as
well. In this case,  
free traffic downstream of the on-ramp is at or below (static) capacity, and
therefore metastable (cf. 
Fig.~\ref{fig:stabdiagIDM}). Consequently, a sufficiently strong perturbation 
is necessary to trigger the WSP, PLC, or
OCT states. Specifically, for the WSP, PLC, and Pinch-OCT simulations of
Fig.~\ref{fig:phasediagIDM2a}, the perturbations associated with
 the mergings at the ramp are not strong enough and an external
perturbation (a moving jam) is necessary to trigger the congested states.
}

In contrast to the GKT simulations, however, the HCT state is obviously
missing. Even for the maximum ramp flows, where merging is
possible for all vehicles (about \unit[1000]{vehicles/h}), the
congested state behind the on-ramp remains oscillatory. This is
consistent with the corresponding stability diagram in
Fig.~\ref{fig:stabdiagIDM}, which shows no restabilization of traffic
flows at high densities, i.e. the critical densities $\rho_3$ and
$\rho_4$ do not exist. This finding, however,
depends on the parameters. It can be analytically 
shown~\citep{Helb-Phases-EPJB-09} that a HCT
state exists, if
\begin{equation}
\label{IDM-HCT}
s_0 < aT^2.
\end{equation}
This means, when varying the minimum distance $s_0$ and leaving all other
IDM parameters constant (at the values given above), a phase diagram of the type shown
in Fig.~\ref{fig:phasediagGKT2b}  (containing oscillatory and homogeneous 
congested traffic patterns) exists for  $s_0 \le \unit[1.2]{m}$, while a phase diagram as in
Fig.~\ref{fig:phasediagIDM2a} (without restabilization at high densities) results for $s_0>\unit[1.2]{m}$.
Moreover, when varying the acceleration parameter $a$ and leaving all other
IDM parameters at the values given above, the IDM phase diagram is of the
type displayed in Fig.~\ref{fig:phasediagIDM2a}, if $\unit[0.93]{m/s^2} \le a <
\unit[1.3]{m/s^2}$, but of the type shown in the original phase
diagram by~\cite{Opus} (without a WSP state), if $a<\unit[0.93]{m/s^2}$, 
and of the type belonging to a single-phase model (with homogeneous traffic states
only), if $a>\unit[1.3]{m/s^2}$.

We obtain the surprising result that, in contrast to the IDM
parameters chosen by~\cite{Opus}, homogeneous congested traffic of
the WSP type can be observed even for very \textit{small}
bottleneck strengths, while the pinch effect is observed for intermediate
bottleneck strengths and a sufficiently high inflow on the freeway
into the bottleneck area. Furthermore, no restabilization takes place for strong
bottlenecks, in agreement with what is demanded by \cite{kerner_jpa_2008}. 
Notice that the empirically observed oscillations are not perfectly periodic as in
Fig.~\ref{fig:phasediagIDM2a}, but quasi-periodic with a continuum of
elementary frequencies concentrated around a typical frequency
(corresponding to a period of about \unit[3.5]{min} in
the latter reference). In computer simulations, such a quasi-periodicity is obtained 
for heterogeneous driver-vehicle units with varying time gaps. 

\begin{figure}
\centering
\includegraphics[width=0.85\textwidth]{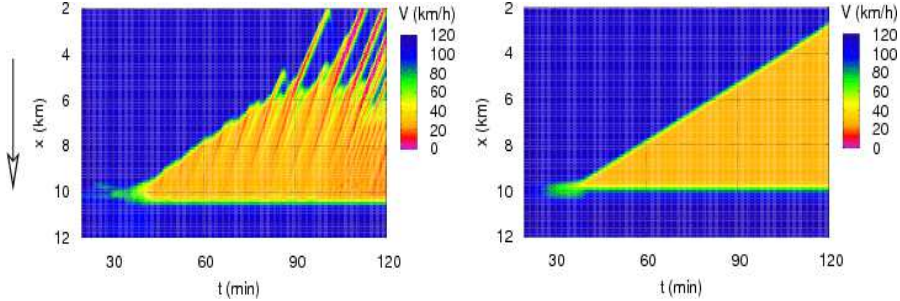}

 \caption{\label{fig:flowcons}Congestion patterns caused (a) by an
 on-ramp bottleneck, and (b) by a comparable flow-conserving
 bottleneck (resulting in the same average traffic flow in the
 congested region). The simulations were performed with the IDM,
 using the same parameters as specified in the main text before.}
\end{figure}
%
Clearly, the merging rule generates considerable noise at the on-ramp. It is therefore
instructive to compare the on-ramp scenario with a scenario assuming a flow-conserving bottleneck, 
but the same model and the same parameters. Therefore, it is 
instructive to simulate a flow-conserving bottleneck rather than an on-ramp bottleneck.
Formally, we have implemented the flow-conserving
bottleneck by gradually increasing the time gap  $T$ from
\unit[1.0]{s} to a higher value $T_{\rm bottl}$ within a
\unit[600]{m} long region as in~\cite{Opus}, keeping
$T=T_{\rm bottl}$ further downstream. The value of
$T_{\rm bottl}$ determines the effectively resulting
\textit{bottleneck strength}. We measure the
bottleneck strength as the difference of 
the outflow from wide moving jams sufficiently away from the 
bottleneck and the average flow in the congested
area upstream of it~\citep{Treiber-TGF99}.

Performing exactly the same simulations as in Fig.~\ref{fig:phasediagIDM2a}, but
replacing the on-ramp bottleneck by a flow-conserving bottleneck, we
find essentially no difference for most combinations of the main
inflow and the bottleneck strength. However, a considerable fraction
of the parameter space leading to a pinch effect in the on-ramp
system results in a WSP state in the case of the flow-conserving bottleneck.
Figure~\ref{fig:flowcons} shows the direct comparison for a main
inflow of $Q_{\rm in}=\unit[2000]{vehicles/h}$, and a ramp flow of
$Q_{\rm rmp}=\unit[250]{vehicles/h}$, corresponding to
$T_{\rm bottl}=\unit[1.37]{s}$ in the flow-conserving system.
It is obvious that nonstationary perturbations are necessary to
trigger the pinch effect, which agrees with the findings  for the GKT
model.

Complementary, we have also investigated other car-following models
such as the model of~\cite{Gipps81}, the optimal velocity model (OVM)
of~\cite{Bando-jphys}, and the velocity difference model (VDM) investigated
by~\cite{Jiang-vDiff01}. We have found that the Gipps model always
produces phase diagrams of the type shown in
Figs.~\ref{fig:phasediagGKT2b} and~\ref{fig:phasediagIDM2a} (see
Fig.~\ref{fig:Gipps} below for a plot of the pinch effect). With the other two
models, it is possible to simulate {\it both} types of diagrams, when the model 
parameters are suitably chosen.

To summarize our simulation results, we have found that the pinch
effect can be produced with two-phase models with particular parameter
choices. Furthermore, nonstationary perturbations clearly favour
the emergence of the pinch effect. In practise, they can originate from
lane-changing maneuvers close to on-ramps, thereby favouring the pinch
effect at on-ramp bottlenecks, while it is less likely to occur at flow-conserving 
bottlenecks. Additionally, nonstationary perturbations can result from noise terms
which are an integral part of essentially all three-phase models
proposed to date.

\begin{figure}
\centering
\includegraphics[width=0.85\textwidth]{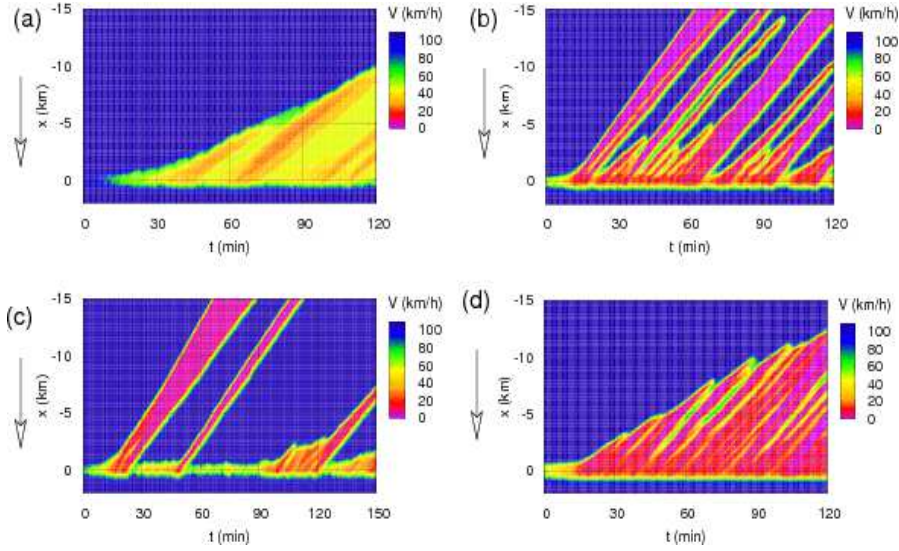}

 \caption{\label{fig:statesKK}Traffic patterns produced by the KK
 model in the open system with an on-ramp (merging length
 \unit[600]{m}). (a) Inflow $Q_{\rm in}=\unit[2100]{vehicles/h}$ and
 ramp flow $Q_{\rm rmp}=\unit[150]{vehicles/h}$; (b)
 $Q_{\rm in}=\unit[2050]{vehicles/h}$,
 $Q_{\rm rmp}=\unit[550]{vehicles/h}$ (c)
 $Q_{\rm in}=\unit[2250]{vehicles/h}$,
 $Q_{\rm rmp}=\unit[320]{vehicles/h}$, and (d)
 $Q_{\rm in}=\unit[1350]{vehicles/h}$,
 $Q_{\rm rmp}=\unit[750]{vehicles/h}$.}

\end{figure}

\begin{figure}
\centering
\includegraphics[width=0.85\textwidth]{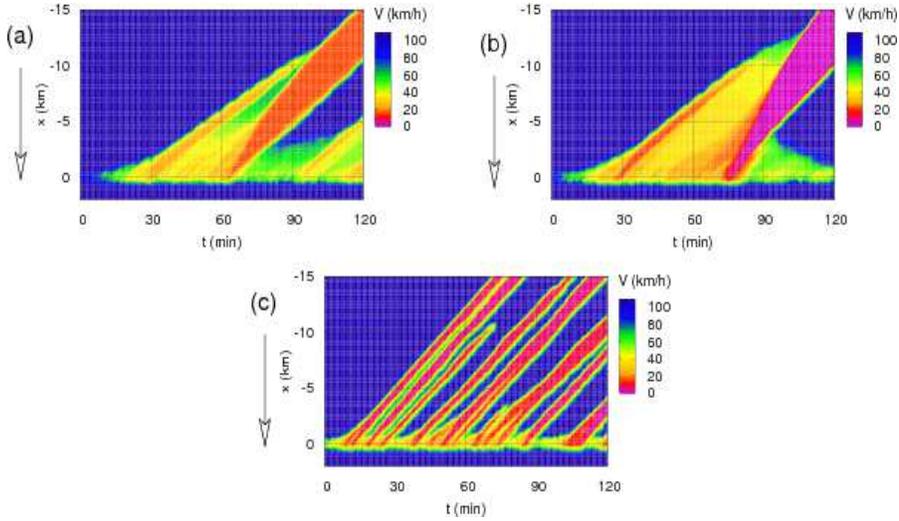}

 \caption{\label{fig:statesKDGP}(a) ``Moving synchronized
 pattern'' (for $Q_{\rm in}=\unit[2120]{vehicles/h}$ and
 $Q_{\rm rmp}=\unit[200]{vehicles/h}$), and (b) ``dissolving
 general pattern'' (for $Q_{\rm in}=\unit[2150]{vehicles/h}$ and
 $Q_{\rm rmp}=\unit[250]{vehicles/h}$), simulated with the KK
 micro-model. The plot (c) shows the pinch effect for 
 $Q_{\rm in}=\unit[1950]{vehicles/h}$, $Q_{\rm rmp}=\unit[500]{vehicles/h}$,
 and the synchronization distance parameter $k=1$, 
 for which the KK micro-model is reduced to a model with a
 unique fundamental diagram.}

\end{figure}

\subsection{\label{sec:three}Three-phase models}

To facilitate a direct comparison of two- and three-phase models, 
we have simulated the same traffic system with two models 
implementing three-phase traffic theory, namely
the cellular automaton of \cite{Kerner-book} and the continuous-in
space model proposed by~\cite{Kerner-Mic}. In the following, we will
focus on the continuous model and refer to it as 
{\it KK micro-model}. It is formulated in terms of a coupled iterated map, i.e.,
the locations and velocities of the vehicles are continuous, but the
updates of the locations and velocities occur in discrete time steps.

To calculate one longitudinal velocity update, 19 update rules have to
be applied (see~\cite{Kerner-book}, Eqs.~(16.41), (16.44)-(16.48),
and the 13 equations of Table~16.5 therein). Besides the vehicle
length, the KK micro-model has 11 parameters and two functions containing
five more constants: The desired velocity $v_{\rm free}$, the time
$\tau$ which represents both, the update time step and the minimum time gap, the
maximum acceleration $a$, the deceleration $b$ for determining the
``safe'' velocity, the synchronization range parameter $k$ indicating
the ratio between maximum and minimum synchronized flow under stationary
conditions at a certain density, the dimensionless sensitivity
$\phi_0$ with respect to velocity differences, a threshold
acceleration $\delta$ that defines, whether the vehicle is in the state
of ``nearly constant speed'', and three probabilities $p_1$, $p_a$, and
$p_b$ defining acceleration noise and a slow-to-start
rule. Additionally, the stochastic part of the model contains the two
probability functions $p_0(v)=0.575+0.125 \min(1, 0.1 v)$ (with $v$ in
units of m/s), and $p_2(v)=0.48$, if $v\le \unit[15]{m/s}$,
otherwise $p_2(v)=0.8$.  The KK micro-model includes further rules for lane
changes and merges.

We have implemented the longitudinal update rules according to the
formulation in~\cite{Kerner-book}, Section~16.3, and used the
parameters from this reference as well. Since we are interested in the
longitudinal dynamics, we will use the simpler merging rule
applied already to the IDM in Sec.~\ref{sec:simIDM} of this paper. 
To test the implementation, we have simulated the open on-ramp system
with a merging length of \unit[600]{m}, as in the other simulations. This  
essentially produced the phase diagram and traffic patterns depicted
in Fig.~18.1 of \cite{Kerner-book}. (Due to the simplified merging rule
assumed here, the agreement is good, but not exact.)

Figure~\ref{fig:statesKK} shows the patterns which are crucial to
compare the KK micro-model with the two-phase models of the previous
section. We observe that the WSP pattern (diagram~(a)), the pinch effect
(diagram~(b)), and the OCT (diagram~(d)) are essentially equivalent with those
of the IDM (Fig.~\ref{fig:phasediagIDM2a}) or the GKT model for parameter 
set~2 (see Fig.~\ref{fig:phasediagGKT2b}), but with the exception of 
the missing HCT states.  Furthermore, the pattern shown in diagram (c) 
resembles the triggered stop-and-go traffic (TSG) displayed in 
Fig.~\ref{fig:phasediagGKT1b}(b). Some differences, however, remain:

\begin{itemize}

\item The oscillation frequencies of oscillatory patterns of the KK
micro-model are smaller than those of the IDM, and often closer to
reality. However, generalizing the IDM by considering reactions 
to next-nearest neighbors~\citep{VDT} increases the
frequencies occurring in the IDM to realistic values. Note that the dynamics in
the KK micro-model depends on next-nearest vehicles as well, so this may be an
important aspect for microscopic traffic models to be realistic.

\item
  The ``moving synchronized patterns'' in Fig.~\ref{fig:statesKDGP}(a)
  (see also Fig.~18.1(d) in~\cite{Kerner-book}) differ from all other
  patterns in that their downstream fronts (where vehicles leave the
  jams) and the internal structures within the congested state
  propagate upstream at {\it different} velocities. Within the KK micro-model,
  the propagation velocity of structures in congested traffic may even exceed
  \unit[40]{km/h} (see, for example, Fig.~18.27 in~\citep{Kerner-book}),
  while there is no empirical evidence of this. Observations 
  rather suggest that the downstream front of congestion patterns 
  is either stationary or propagates at a characteristic speed  
  (see {\it stylized fact 4} in Sec.~\ref{sec:3ddata}). 

\item Another pattern which is sometimes produced by three-phase models
is the ``dissolving general pattern'' (DGP), where an emerging wide moving jam leads 
to the dissolution of  synchronized traffic (Fig.~\ref{fig:statesKDGP}(b), see also
Fig.~18.1(g) in ~\cite{Kerner-book}). So far, we have not found any evidence for such a 
pattern in our extensive empirical data sets. Congested traffic normally dissolves
in different ways (see {\it stylized facts 4 and 5}).
\end{itemize}

Finally, we observe that the time gap $T$ of the KK micro-model in stationary
car-following situations can adopt a range given by $\tau
\le T \le k\tau$, where $k$ is the synchronization distance
factor. By setting $k=1$, the KK micro-model becomes a
conventional two-phase model. When simulating the on-ramp scenario
for the KK micro-model with $k=1$, we essentially found the same patterns
(see Fig.~\ref{fig:statesKDGP}(c) for an example). This suggests that 
there is actually no need of going beyond the simpler class of two-phase models with
a unique fundamental diagram.

\subsection{\label{sec:pinch}Different mechanisms producing the pinch effect}

While the very first publications on the phase diagram of traffic states 
did not report a pinch effect (or ``general pattern''), 
the previous sections of this paper have shown that this 
traffic pattern {\it can} be simulated by two-phase models,
if the model parameters are suitably chosen. It also appears that
nonstationary perturbations at a bottleneck (which may, for example,
result from frequent lane changes due to weaving flows) support the 
occurrence of a pinch effect. This suggests to take a closer look
at mechanisms, which produce this effect. We have identified three
possible explanations, which are discussed in the following. In reality, one may 
also have a combination of these mechanisms.

\subsubsection{\label{sec:GP1}Mechanism I: 
metastability and depletion effect}

This mechanism is the one proposed by three-phase traffic theory. The
starting point is a region with metastable congested (but flowing) 
traffic behind a bottleneck, while sufficiently large
perturbations trigger small oscillations in the density or velocity that
grow while propagating upstream.  When they become fully developed
jams, the outflow from the oscillations decreases, which is modeled by some sort
of \textit{slow-to-start rule}: Once stopped or forced to drive at
very low velocity, drivers accelerate more slowly, or keep a longer
time gap than they would do when driving at a higher velocity. In the
KK micro-model, this effect is implemented by using velocity-dependent
stochastic deceleration probabilities $p_0(v)$ and $p_2(v)$.  Other
implementations of this effect are possible as well, such as the
memory effect~\citep{IDMM}, or a driving style that depends on the
local velocity variance~\citep{VDT}. Even the parameters $s_0$ and
$s_1$ of the IDM can be used to reflect this effect.

In any case, as soon as the outflow from large jams becomes smaller
than that from small jams, most of the latter will eventually dissolve,
resulting in only a few ``wide moving jams''. We call this the
\textit{``depletion effect''}.

\subsubsection{\label{sec:GP2}Mechanism II: 
convective string instability}

A typical feature of the pinch effect are small perturbations
that grow to fully developed moving jams. Therefore, it is expected that
(linear or nonlinear) instabilities of the traffic flow play an essential
role. However, another characteristic feature of the pinch effect is a
stationary congested region near the bottleneck, called the
\textit{pinch region}~\citep{KernerPinch}.

The simultaneous observation of the stationary pinch region and
growing perturbations upstream of it can be naturally explained by
observing that, in spatially extended open systems (such as traffic
systems), there are two different types of string
instability~\citep{huerre1990lag,ThreeTimes-07}. For the first type, an
\textit{absolute instability}, the perturbations will eventually
spread over the whole system. A pinch region, however, can only
exist if the growing perturbations propagate away from the on-ramp
(in the upstream direction), while they do not ``infect'' the bottleneck
region itself. This corresponds to the second type of string
instability called \textit{``convective instability''}.

\begin{figure}
\centering
\includegraphics[width=0.5\textwidth]{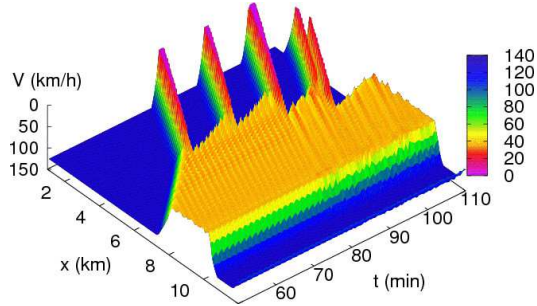}

 \caption{\label{fig:Gipps}Simulation of the on-ramp system with the
 Gipps model showing the pinch effect. The parameters $v_0$, $a$, $b$
 of this model have the same meaning as for the IDM and 
 have been set to the same values (see main text). The update
 time $\Delta t$ (playing also the role of the time gap) has been set
 to $\Delta t=\unit[1.2]{s}$.}

\end{figure}

Figure~\ref{fig:Gipps} illustrates convectively unstable traffic by a
simulation of the bottleneck system with the model of~\cite{Gipps81}:
Small perturbations caused by the merging maneuvers near the on-ramp
at $x=\unit[10]{km}$ grow only in the upstream direction and
eventually transform to wide jams a few kilometers upstream. The IDM
simulations of Fig.~\ref{fig:phasediagIDM2a} show this mechanism as
well.

The concept of convective instability, which has been introduced into the context of traffic modeling
already some years ago~\citep{Phase}, is in agreement empirical evidence.
It has been observed that, in extended congested traffic, small perturbations or oscillations may grow while propagating
upstream, whereas congested traffic is relatively stationary in the vicinity of the bottleneck
\citep{Kerner-book,Mauch-Cassidy,DaganzoLong1999,Zielke-intlComparison}. 
However, the congestion pattern emanating from the ``pinch
region'' is not necessarily a fully developed ``general pattern'' in
the sense that it includes a pinch
region, small jams, and a transition to wide jams \citep{Kerner-book}. In fact, the pinch
region is also observed as part of congestion patterns that include
neither wide jams nor a significant number of merging events, see
Fig.~\ref{fig:empdata1}(a) for an example.
This can be understood by assuming that the
mechanisms leading to the pinch region and to wide jams are
essentially independent from each other. One could therefore
explain the pinch region by the convective instability, and the
transition from small to wide jams by the depletion effect (see
Sec.~\ref{sec:GP1}).
%

\subsubsection{Mechanism III: locally increased stability}

\begin{figure}
\centering

 \includegraphics[width=0.8\textwidth]{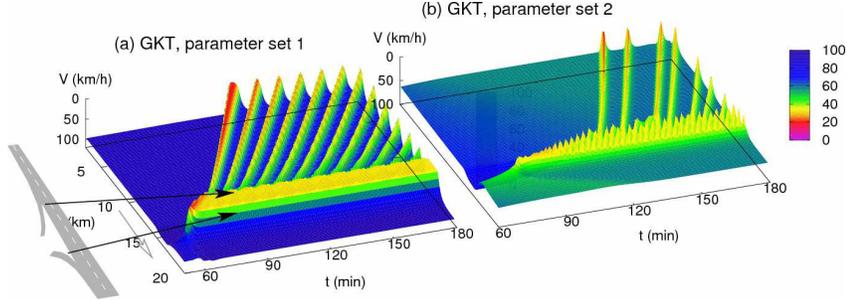}
 
 \caption{\label{fig:pinchOffOn}Congested traffic at a combination of an off-ramp
 with an on-ramp, simulated with
 the GKT model. The locally increased stability between the ramps
supports the pinch and depletion effects, leading to a 
composite pattern consisting of a pinch region, 
narrow jams, and wide moving jams (see main text for details). 
}

\end{figure}

A third mechanism leading to similar results as the previous
mechanisms comes into play at intersections and junctions, where
off-ramps are located upstream of on-ramps (which corresponds to the
usual freeway design). Figure~\ref{fig:pinchOffOn}(a) illustrates this
mechanism for the GKT model and the parameter set~1 in
Table~\ref{tab:GKT}. The existence of a stationary and essentially
homogeneous pinch region and a stop-and-go pattern further upstream
can be explained, assuming that the inflow $Q_{\rm rmp,on}$ from the
on-ramp (located downstream) must be sufficiently large such that a
HCT or OCT state would be produced when simulating this on-ramp
alone. Furthermore, the outflow $Q_{\rm rmp,off}$ from the off-ramp
must be such that an \textit{effective} on-ramp of inflow
\begin{equation}
\label{onoff}
Q_{\rm rmp,eff}=Q_{\rm rmp,on}-Q_{\rm rmp,off}
\end{equation}
would produce a TSG state or an OCT state with a larger wavelength. 

Figure~\ref{fig:pinchOffOn}(b) shows a simulation of an
off-ramp-on-ramp scenario with the GKT model and parameter
set~2 in Table~\ref{tab:GKT}. 
Notice that, for the parameters chosen, 
a pinch effect is \textit{not} possible at an {\it isolated} on-ramp without a 
previous off-ramp (see Fig.~\ref{fig:phasediagGKT1b}).
In Fig.~\ref{fig:pinchOffOn}(b), the on-ramp produces 
an OCT pattern, and the effective ramp flows according to Eq.~\eqref{onoff} implies TSG
traffic (or OCT with larger oscillation periods). The difference between the
oscillation periods of the congestion pattern upstream of the on-ramp
and upstream of the off-ramp leads to merging phenomena which are 
similar to those caused by the depletion effect. Notice that the existence of the
depletion effect in congestion patterns forming behind intersections 
depends not only on the chosen traffic model and its
parameters, but also on the traffic volume and the intersection
design.  This is in agreement with observations showing that some
intersections tend to produce the full composite pattern consisting of
the pinch region, narrow jams, and wide jams, while wide moving jams  
are missing at others~\citep{Kerner-book,Martin-empStates}.
Furthermore, a pinch effect is usually not observed at flow-conserving bottlenecks
\citep{Martin-empStates}, and often not at separated on-ramps (see the traffic video at http://www.trafficforum.org/stopandgo).

\providecommand{\entry}[3]{
       \hspace{0mm}\parbox{35mm}{\vspace*{1.2mm}#1
          \vspace*{1.2mm}} \hspace*{-2mm}
     & \hspace{0mm}\parbox{30mm}{\vspace*{1.2mm}#2
         \vspace*{1.2mm}} \hspace*{-2mm}
     & \hspace{0mm}\parbox{65mm}{\vspace*{1.2mm}#3
         \vspace*{1.2mm}} \hspace*{-2mm}
         \\  
}

\begin{table}

 \caption{\label{tab:mech}Overview of the main controversial traffic
 phenomena and their possible explanations. The term ``three-phase
 model'' has been used for models that are consistent with Kerner's
 theory, while two-phase models are conventional models that can
 display traffic instabilities such as second-order macroscopic models
 and most car-following models (see Sec. \protect\ref{sec:phase} for details).}

\begin{center}

\begin{tabular}{|c|c|c|}\hline
\entry{Phenomenon}{Possible\\Mechanism}{Examples and Models} \hline\hline
\entry{Pinch region at a\\ bottleneck;\\small jams
further upstream}{1. Convective
in\-sta\-bi\-li\-ty or meta\-sta\-bi\-li\-ty}{
 (i) Three-phase models\\
(ii) Two-phase models with appropriate parameters}
\entry{}{2. Local change\\ of stability\\and capacity}{Off-ramp-on-ramp combinations}
\hline
\entry{Transition from\\small to wide jams}{1. Depletion\\mechanism}
{Slow-to-start rule and other forms of intra-driver variability}
\entry{}{2. Merging\\mechanism}
{Different group velocities of the small waves}
\hline
\entry{Homogeneous\\con\-ges\-ted traffic at low densities}
{Maximum flow is\\metastable or stable}
{Two- and three-phase models with suitable
parameters}
\hline
\entry{Homogeneous\\con\-ges\-ted traffic at high densities}{Restabilization}
{Severe bottleneck simulated with a two-phase model with appropriate
parameters} 
\hline
\end{tabular}
\end{center}

\end{table}

\subsection{\label{sec:scatter}Summary of possible explanations}

Table~\ref{tab:mech} gives an overview of mechanisms 
producing the observed spatiotemporal phenomena listed in Sec.~\ref{sec:3ddata}.
So far, these have been either considered incompatible with three-phase models or
with two-phase models having a fundamental diagram. It is
remarkable that the main controversial observation --- the occurrence of the pinch effect
or general pattern --- is not only compatible with three-phase models,
but can also be produced with conventional two-phase models. For both model 
classes, this can be demonstrated with macroscopic, microscopic, and cellular automata
models, if models and parameters are suitably chosen.

\section{\label{sec:conclusions}Conclusions}

It appears that some of the current controversy in the area of traffic modeling
arises from the different
definitions of what constitutes a traffic phase. In the context of
three-phase traffic theory, the definition of a phase is oriented at
equilibrium physics, and in principle, it should be able to determine 
the phase based on {\it local} criteria and measurements at a {\it single}
detector. Within three-phase traffic theory, however, this goal is not completely 
reached: In order to distinguish
between ``moving synchronized patterns'' and wide moving jams, which look
alike, one needs  the additional \textit{nonlocal} criterium of whether the congestion
pattern propagates through the \textit{next} bottleneck area or
not~\citep{Martin-empStates,Schoenhof-TRB09}.  In contrast,
the alternative phase diagram approach is oriented at systems theory,
where one tries to distinguish different kinds of elementary 
congestion patterns, which may be considered as non-equilibrium phases
occurring in non-homogeneous systems (containing bottlenecks). These
traffic patterns are distinguished into localized or spatially extended,
moving or stationary (``pinned''), and spatially homogeneous or oscillatory patterns.
These patterns can be derived from the stability properties of
conventional traffic models exhibiting a unique fundamental diagram and
unstable and/or metastable flows under certain conditions. Models of this
class, sometimes also called \textit{two-phase} models, include macroscopic 
and car-following models as well as cellular automata.

As key result of our paper we have found that features, which are 
claimed to be consistent with three-phase traffic theory
only, can also be explained and simulated with conventional
models, if the model parameters are suitably specified. In particular, 
if the parameters are
chosen such that traffic at maximum flow is (meta-)stable and the
density range for unstable traffic lies completely on the
``congested'' side of the fundamental diagram, we find the
``widening synchronized pattern'' (WSP), which has not been discovered in 
two-phase models before. Furthermore, the models can be tuned such that no
homogeneous congested traffic (HCT) exists for strong bottlenecks. 
Conversely, we have shown that almost the same kinds of patterns, which are
produced by two-phase models, are also found for models 
developed to reproduce three-phase traffic theory (such as the KK micro-model). 
Moreover, when the KK micro-model is simulated with
parameters for which it turns into a model with a unique fundamental diagram, 
it still displays very similar results. Therefore, the difference between
so-called two-phase and three-phase models does not seem to be as big 
as the current scientific controversy suggests.

For many empirical observations, we have found {\it several} plausible
explanations (compatible and incompatible ones), which makes it difficult to 
determine the underlying mechanism 
which is actually at work. In our opinion, convective instability is a likely reason
for the occurence of the pinch effect (or the general pattern), but
at intersections with large ramp flows, the effect of off- and on-ramp combinations
seems to dominate. To explain the transition to wide moving jams, we favour
the depletion effect, as the group
velocities of structures within congested traffic patterns are essentially
constant. For the wide scattering of flow-density data, all three
mechanisms of Table~\ref{tab:mech} do probably play a role.  Clearly,
further observations and experiments are necessary to confirm or
reject these interpretations, and to exclude some of the alternative
explanations. It seems to be an interesting challenge for the future
to devise and perform suitable experiments in order to finally decide between the alternative 
explanation mechanisms.
\vspace{1em}

In our opinion, the different congestion patterns produced by three-phase traffic theory 
and the alternative phase diagram approach for models with a 
fundamental diagram share more commonalities than differences. Moreover, according
to our judgement, three-phase models do not explain {\it more} observations than
the simpler two-phase models (apart maybe from the fluctuations of ``synchronized flow'', which can, for example, be explained by the heterogeneity of driver-vehicle units). The question is, therefore, which approach is superior over the other. To decide this,
the quality of models should be judged in a {\it quantitative} way, applying the following established
standard procedure \citep{Greene,Diebold}:
\begin{itemize}
\item As a first step, mathematical quality functions must be defined. Note that the proper selection 
of these functions (and the relative weight that is given to them) depends on the 
purpose of the model.\footnote{For example, travel times may be the most relevant quantity for
traffic forecasts, and macroscopic models or extrapolation models may be good
enough to provide reasonably accurate results at low costs. However, if the impact of
driver assistance systems on traffic flows is to be assessed, it is important to accurately
reproduce the time-dependent speeds, distances, and accelerations as well,
which calls for microscopic traffic models.}
\item The crucial step is the statistical comparison of the competing models based on a new,
but representative set of traffic measurements, using model parameters determined
in a previous calibration step. Note that,  due to the problem of 
over-fitting (i.e. the risk of fitting of noise in the data), a high goodness of fit in
the calibration step does not necessarily imply a good fit of the new data set, 
i.e. a high predictive power \citep{Wagner2,Wagner1}.
\item The goodness of fit should be judged with established statistical methods, for example with the adjusted R-value or similar concepts considering the number of model parameters \citep{Greene,Diebold}.
Given the same correlation with the data, a model containing a few parameters 
has a higher explanatory power than a model with many parameters. 
\end{itemize}
Given a comparable predictive power of two models, one should select 
the simpler one according to Einstein's principle that
\textit{a model should be as simple as possible, but not
simpler}. If one has to choose between two equally performing models 
with the same number of parameters, one should use the one which is
easier to interpret, i.e. a model with meaningful and independently 
measurable parameters (rather than just fit parameters). Furthermore,
the model should not be sensitive to variations of the model parameters within 
the bounds of their confidence intervals. Applying this benchmarking process to traffic 
modeling will hopefully lead to an eventual convergence of explanatory concepts in traffic theory.

\subsection*{Acknowledgements}
The authors would like to thank the {\it Hessisches Landesamt f\"ur
Stra{\ss}en- und Verkehrswesen} and the {\it Autobahndirektion
S\"udbayern} for providing the freeway data shown in Figs. 1 and 2. They are furthermore grateful
to Eddie Wilson for sharing the data set shown in Fig. \ref{fig:Wilson}, and to Anders Johansson
for generating the plots from his data.

\appendix

\section{Wide scattering of congested flow--density data}\label{SCA}

The discussion around three-phase traffic theory is directly related with the wide
scattering of flow-density data within synchronized traffic flows.
However, it deserves to be mentioned that the discussion
around traffic theories has largely neglected the fact that empirical
measurements of wide moving jams show a considerable amount of
scattering as well (see, e.g. Fig. 15 of~\cite{Opus}), while theoretically, one expects
to find a ``jam line''~\citep{Kerner-book}. This suggests
that wide scattering is actually {\it not} a specific feature of
synchronized flow, but of congested traffic in general. While this
questions the basis of three-phase traffic theory to a certain extent,
particularly as it is claimed that wide scattering is a distinguishing
feature of synchronized flows as compared to wide moving jams, the
related car-following models \citep{Kerner-Mic}, cellular
automata \citep{KKW-CA,KKWimproved}, and
macroscopic models \citep{jiang2007-threePhaseMacro} build in
dynamical mechanisms generating such scattering as one of their key
features~\citep{siebel2006}. In other models, particularly those
with a fundamental diagram, this scattering is a simple add-on (and
partly a side effect of the measurement process, see
Sec.~\ref{sec:measurement}). It can be reproduced, for example, by considering
heterogeneous driver-vehicle populations in macroscopic
models~\citep{Wagner-96,Krauss-Wagner,Banks-scatter,GKT-scatter,hoogendoorn2000-multiclass}
or car-following
models~\citep{Katsu03,Ossen-interDriver06,Igerashi-interIntraDriver,Kesting-Calibration-TRR08},
by noise terms~\citep{VDT}, or slowly changing driving
styles~\citep{IDMM,HDM}.

\par
\begin{figure}
\centering

 \includegraphics[width=0.6\textwidth]{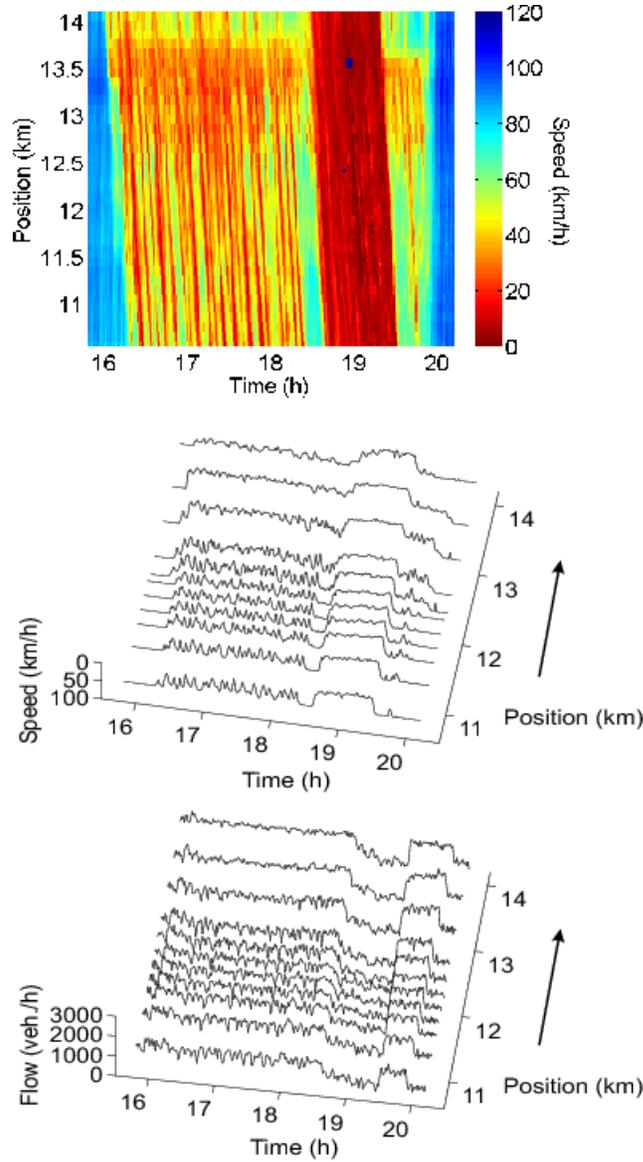}

 \caption[]{Homogeneous congested traffic on the high-coverage section
of the British freeway M42 ATM (averaged over 3 running lanes)
\protect\citep{Wilson}. Note that no interpolation or smoothing was
applied to the data measured on November 27, 2008. The
three-dimensional plots of the vehicle speed and the flow show
measurements of each fifth detector only, otherwise the plots would
have been overloaded. There is no clear evidence that perturbations in
the vehicle speed or flow would grow upstream, i.e. against the flow
direction that is indicated by arrows.} 
 \label{fig:Wilson}
\end{figure}

\section{Discussion of homogeneous congested traffic}\label{APP}

For strong bottlenecks (typically caused by accidents), empirical evidence regarding the existence of
homogeneous congested traffic has been somewhat ambiguous so far. On the one hand, when applying the adaptive smoothing method to get rid of noise in the data~\citep{Treiber-smooth}, the spatiotemporal speed profile looks almost homogeneous,  even when the same smoothing parameters are used as for the measurement of the other traffic patterns, e.g. oscillatory ones \citep{Martin-empStates}. On the other hand, it was claimed that data of the flow measured at freeway cross sections show an oscillatory behavior \citep{kerner_jpa_2008}. These oscillations typically have small wavelengths, which can have
various origins: (1) They can result from the heterogeneity of
driver-vehicle units, particularly their time gaps, which is known to
cause a wide scattering of congested flow-density data \citep{Katsu03}. 
(2) They could as well result from problems in
maintaining low speeds, as the gas and break pedals are difficult to
control. (3) They may also be a consequence of perturbations, which can
easily occur when traffic flows of several lanes have to merge in a
single lane, as it is usually the case at strong
bottlenecks. According to {\it stylized fact 6}, all these
perturbations are expected to propagate upstream at the speed $c_{\rm
cong}$. In order to judge whether the pattern is to be classified as
oscillatory congested traffic or homogeneous congested traffic, one
would have to determine the sign of the growth rate of perturbations,
i.e. whether large perturbations grow bigger or smaller while
travelling upstream. 

Recent traffic data of high spatial and temporal resolution suggest that 
homogeneous congested traffic states {\it do} exist (see Fig. \ref{fig:Wilson}), but are very rare. 
For the conclusions of this paper and the applicability of the phase diagram approach, however, it does not 
matter whether homogeneous congested traffic actually exists or
not. This is, because many models with a fundamental diagram can be
calibrated in a way that either generates homogeneous patterns for
high bottleneck strengths or not (see Sec. \ref{sec:phase}).  

\clearpage




\end{document}